\documentclass[12pt,letter]{article}

\usepackage{graphicx, epsfig, color, cite}
\def\eslt{E_T^{\rm miss}}
\def\to{\rightarrow}

\def\bi{\begin{itemize}}
\def\ei{\end{itemize}}

\def\ta{\tilde a}

\def\tst{\tilde t}

\def\tg{\tilde g}

\def\tq{\tilde q}
\def\tw{\widetilde\chi^{\pm}}
\def\tz{\widetilde\chi^0}

\def\agt{\stackrel{>}{\sim}}
\def\be{\begin{equation}}  
\def\ee{\end{equation}}  

\newcommand\prd[3]{{\it Phys.\ Rev.\ }{\bf D #1} (#2) #3}

\newcommand\prl[3]{{\it Phys.\ Rev.\ Lett.\ }{\bf #1} (#2) #3}
\newcommand\plb[3]{{\it Phys.\ Lett.\ }{\bf B #1} (#2) #3}

\newcommand\jhep[3]{{\it J. High Energy Phys.\ }{\bf #1} (#2) #3}

\newcommand\npb[3]{{\it Nucl.\ Phys.\ }{\bf B #1} (#2) #3}
\newcommand\epjc[3]{{\it Eur.\ Phys.\ J. }{\bf C #1} (#2) #3}
\newcommand\cpc[3]{{\it Comput. Phys. Commun.}{\bf #1} (#2) #3}
\newcommand\ptp[3]{{\it Prog.\ Theor.\ Phys.\ }{\bf #1} (#2) #3}

\newcommand{\hepph}[1]{hep-ph/#1}

\newcommand\arnps[3]{{\it Ann.\ Rev.\ Nucl.\ Part.\ Sci.}{\bf  #1} (#2) #3}

\begin{document}
\begin{titlepage}

\begin{flushright}
OU-HEP/091015\\
LPSC09155
\end{flushright}

\vspace*{0.5cm}
\begin{center}
{\Large \bf 
Beyond the Higgs boson at the Tevatron:\\[2mm]
detecting gluinos from Yukawa-unified SUSY
}\\
\vspace{1.2cm} \renewcommand{\thefootnote}{\fnsymbol{footnote}}
{\large Howard Baer$^{1}$\footnote[1]{Email: baer@nhn.ou.edu },
Sabine Kraml$^{2}$\footnote[2]{Email: sabine.kraml@lpsc.in2p3.fr},
Andre Lessa$^{1}$\footnote[3]{Email: lessa@nhn.ou.edu},\\
Sezen Sekmen$^3$\footnote[4]{Email: sezen.sekmen@cern.ch}
and Heaya Summy$^1$\footnote[5]{Email: heaya@nhn.ou.edu }} \\
\vspace{1.2cm} \renewcommand{\thefootnote}{\arabic{footnote}}
{\it 
$^1$Dept. of Physics and Astronomy,
University of Oklahoma, Norman, OK 73019, USA \\
$^2$Laboratoire de Physique Subatomique et de Cosmologie, UJF Grenoble 1, 
CNRS/IN2P3, INPG, 53 Avenue des Martyrs, F-38026 Grenoble, France\\
$^3$Dept.\ of Physics, Florida State University, Tallahassee, FL 32306\\
}

\end{center}

\vspace{0.5cm}
\begin{abstract}
\noindent 
Simple SUSY GUT models based on the gauge group $SO(10)$ 
require $t-b-\tau$ Yukawa coupling unification, in addition 
to gauge coupling and matter unification. The Yukawa coupling unification
places strong constraints on the expected superparticle mass spectrum,
with scalar masses $\sim 10$ TeV while gluino masses are 
much lighter: in the 300--500 GeV range.
The very heavy squarks suppress negative interference 
in the $q\bar{q}\to\tg\tg$
cross section, leading to a large enhancement in  production rates.
The gluinos decay almost always via three-body modes into a pair
of $b$-quarks, so we expect at least four $b$-jets per signal event.
We investigate the capability of Fermilab Tevatron collider experiments
to detect gluino pair production in Yukawa-unified SUSY. By requiring
events with large missing $E_T$ and $\ge 2$ or 3 tagged $b$-jets, we find
a 5$\sigma$ reach in excess of $m_{\tg}\sim 400$ GeV for 5 fb$^{-1}$ of data.  
This range in $m_{\tg}$ is much further than the conventional 
Tevatron SUSY reach, and should cut a significant swath through 
the most favored region of parameter space for Yukawa-unified SUSY models.

\vspace{0.8cm}

\end{abstract}


\end{titlepage}

\section{Introduction}
\label{sec:intro}

There is now an ongoing huge effort at the Fermilab Tevatron $p\bar{p}$ collider
to extract a Standard Model Higgs boson 
signal from a daunting set of SM background processes. 
While such an effort is to be lauded-- and if successful
would complete the picture provided by the Standard Model (SM)-- 
we note here that an even bigger 
prize may await in the form of the gluino of supersymmetric (SUSY) models\cite{wss}.
Current searches from CDF and D0 collaborations have explored values of 
$m_{\tg}$ up to $\sim 300$ GeV
within the context of the minimal supergravity (mSUGRA or CMSSM) model\cite{cdf,d0}. 
Here, we show that
Tevatron experiments should-- with current data sets-- be able to expand their gluino search much further:
into the 400 GeV regime, in Yukawa-unified SUSY, which is a model with arguably 
much higher motivation than mSUGRA\cite{raby}. 
Since Yukawa-unified SUSY favors a light gluino in the mass range $300-500$ GeV, 
with the lower portion of this range giving the most impressive Yukawa 
coupling unification\cite{abbbft,bkss,agrs,bhkss,dr3},
such a search would explore a huge swath of the expected model parameter space.

Supersymmetric grand unified theories (SUSY GUTs) based upon the gauge group 
$SO(10)$ are extremely compelling\cite{so10}. 
For one, they explain the ad-hoc anomaly cancellation within the 
SM and $SU(5)$ theories. Further, they unify
all matter of a single generation into the 16-dimensional spinor
representation $\hat{\psi}(16)$, provided one adds to the set of supermultiplets
a SM gauge singlet superfield $\hat{N}^c_i$ ($i=1-3$ is a generation index) 
containing a right-handed neutrino.\footnote{Here, we adopt the 
superfield ``hat'' notation as presented in Ref.~\cite{wss}.}
Upon breaking of $SO(10)$, a superpotential term 
$\hat{f}\ni {1\over 2}M_{N_i}\hat{N}^c_i\hat{N}^c_i$ is induced which 
allows for a Majorana neutrino mass $M_{N_i}$ which is necessary for
implementing the see-saw mechanism for neutrino masses\cite{seesaw}. 
In addition, in the simplest $SO(10)$ theories where the MSSM Higgs doublets reside
in a 10 of $SO(10)$, one expects $t-b-\tau$ Yukawa coupling unification in addition 
to gauge coupling unification at scale $Q=M_{\rm GUT}$\cite{early,also}.
In models with Yukawa coupling textures and family symmetries, one only expects
Yukawa coupling unification for the third generation\cite{dr}.

In spite of these impressive successes, GUTs and also SUSY GUTs have been beset
with a variety of problems, most of them arising from implementing GUT 
gauge symmetry breaking via large, unwieldy Higgs representations.
Happily, in recent years physicists have learned that GUT theories--
as formulated in spacetime dimensions greater than four-- can use
extra-dimension compactification to break the GUT symmetry instead\cite{xdguts}. 
This is much in the spirit of string theory, where anyway one must pass
from a 10 or 11 dimensional theory to a 4-d theory via some sort of
compactification. 

Regarding Yukawa coupling unification in $SO(10)$, 
our calculation begins with stipulating the $b$ and $\tau$
running masses at scale $Q=M_Z$ (for two-loop running, we adopt the 
$\overline{DR}$ regularization scheme) and the $t$-quark running 
mass at scale $Q=m_t$. The Yukawa couplings are evolved to scale $Q=M_{\rm SUSY}$, 
where threshold corrections are implemented\cite{hrs}, as we pass from the 
SM effective theory to the Minimal Supersymmetric Standard Model (MSSM) effective theory.
From $M_{\rm SUSY}$ on to $M_{\rm GUT}$, Yukawa coupling evolution is performed using two-loop
MSSM (or MSSM+RHN) RGEs. 
Thus, Yukawa coupling unification ends up depending on the complete SUSY mass spectrum
via the $t$, $b$ and $\tau$ self-energy corrections.

In this work, we adopt the Isajet 7.79 program for calculation of the 
SUSY mass spectrum and mixings\cite{isajet}.
Isajet uses full two-loop RG running for all gauge and
Yukawa couplings and soft SUSY breaking (SSB) terms. In running from $M_{\rm GUT}$ 
down to $M_{weak}$, the RG-improved 1-loop effective potential is minimized at 
an optimized scale choice $Q=\sqrt{m_{\tst_L}m_{\tst_R}}$, which accounts for
leading two-loop terms. Once a tree-level SUSY/Higgs  spectrum is calculated, the
complete 1-loop corrections are calculated for all SUSY/Higgs particle masses.
Since the SUSY spectrum is not known at the beginning of the calculation, an iterative
approach must be implemented, which stops when an appropriate convergence criterion is satisfied.

Yukawa coupling unification has been examined in a number of previous 
papers\cite{early,abbbft,bkss,agrs,bhkss,dr3,also,bdft,bf,bdr}. 
The parameter space to be considered is given by
\be
m_{16},\ m_{10},\ M_D^2,\ m_{1/2},\ A_0,\ \tan\beta ,\ sign (\mu ) 
\label{eq:pspace}
\ee
along with the top quark mass, which we take to be $m_t=172.6$ GeV\cite{mtop}.
Here, $m_{16}$ is the common mass of all matter scalars at $M_{\rm GUT}$, 
$m_{10}$ is the common Higgs soft mass at $M_{\rm GUT}$ and $M_D^2$ 
parameterizes either $D$-term splitting (DT)\cite{bf,abbbft,dr3} or ``just-so'' Higgs-only soft mass splitting (HS)\cite{bdr,abbbft,bkss}.
The latter is given by $m_{H_{u,d}}^2=m_{10}^2\mp 2M_D^2$. 
As in the minimal supergravity (mSUGRA) model, $m_{1/2}$ is a common GUT scale gaugino mass, 
$A_0$ is a common GUT scale trilinear soft term, and the bilinear SSB term $B$ has been traded for 
the weak scale value of $\tan\beta$ via the EWSB minimization conditions. The latter also
determine the magnitude (but not the sign) of the superpotential Higgs mass term $\mu$.

What has been learned is that $t-b-\tau$ Yukawa coupling unification {\it does} occur in the 
MSSM for $\mu >0$ (as preferred by the $(g-2)_\mu$ anomaly), 
but {\it only if certain conditions} are satisfied.
\bi
\item $\tan\beta \sim 50$.
\item The gaugino mass parameter $m_{1/2}$ should be as small as possible.
\item The scalar mass parameter $m_{16}$ should be very heavy: in the range $8-20$ TeV.
\item The SSB terms should be related as $A_0^2=2m_{10}^2=4m_{16}^2$, with
$A_0<0$ (we use SLHA\cite{slha} conventions). This combination
was found to yield a radiatively induced inverted scalar mass hierarchy (IMH) by Bagger
{\it et al.} \cite{bfpz} for MSSM+right hand neutrino (RHN) 
models with Yukawa coupling unification.
\item EWSB can be reconciled with Yukawa unification only if the Higgs SSB masses
are split at $M_{\rm GUT}$ such that $m_{H_u}^2 <m_{H_d}^2$.\footnote{An exception 
is the case of highly split trilinears \cite{splita}.} The HS prescription ends up
working better than DT splitting\cite{bdr,bf}. 
\ei

In the case where the above conditions are satisfied, Yukawa coupling unification to within 
a few percent can be achieved. The resulting sparticle mass spectrum has some notable 
features.
\bi
\item First and second generation matter scalars have masses of order $m_{16}\sim 8-20$ TeV.
\item Third generation scalars, $m_A$ and $\mu$ are suppressed relative to $m_{16}$
by the IMH mechanism: they have masses on the $1-2$ TeV scale. This reduces the amount of 
fine-tuning one might otherwise expect in such models. 
\item Gaugino masses are quite light, with $m_{\tg}\sim 300-500$ GeV, 
$m_{\tz_1}\sim 50-80$ GeV and $m_{\tw_1}\sim 100-160$ GeV.
\ei

Since the lightest neutralino of $SO(10)$ SUSY GUTs is nearly a pure bino state, it turns out that
its relic density $\Omega_{\tz_1}h^2$ would be extremely high, of order $10^2-10^4$
(unless it annihilates resonantly through the light Higgs\cite{bkss}, which is the case only in
a very narrow strip of the parameter space). 
Such high values conflict with the WMAP observation\cite{wmap}, which gives
\be
\Omega_{CDM}h^2\equiv \rho_{CDM}/\rho_c =
0.1099\pm 0.0124\ \ (2\sigma ) .
\label{eq:Oh2}
\ee
where $h=0.74\pm 0.03$ is the scaled Hubble constant.
Several solutions to the $SO(10)$ SUSY GUT dark matter problem have been proposed in
Refs. \cite{abbk,bkss,bhkss}. The arguably most attractive one is that
the dark matter particle is in fact not the neutralino, but instead a mixture of 
{\it axions} $a$ and thermally and non-thermally produced {\it axinos} $\ta$. 
Mixed axion/axino dark matter occurs in models where the MSSM is extended via the Peccei-Quinn (PQ)
solution to the strong $CP$ problem\cite{nr}. The PQ solution introduces a spin-0 axion field
into the model; if the model is supersymmetric, then a spin-${1\over 2}$ axino
is also required.  The 
$SO(10)$ SUSY GUT models with mixed axion/axino DM  can\cite{bhkss} 1.\ yield the correct
abundance of CDM in the universe (where a dominant {\it axion} abundance is most favorable), 
2.\ avoid the gravitino/BBN problem via 
$m(gravitino)\sim m_{16}\sim 10$ TeV and 3.\ have a compelling
mechanism for generating the matter-antimatter asymmmetry of the universe via 
non-thermal leptogenesis\cite{ntlepto}. A consequence of the mixed axion/axino CDM scenario with an
axino as LSP is that WIMP search experiments will find null results, 
while a possible positive result might be found at relic axion search experiments\cite{admx}.

\begin{figure}[t]
\begin{center}
\epsfig{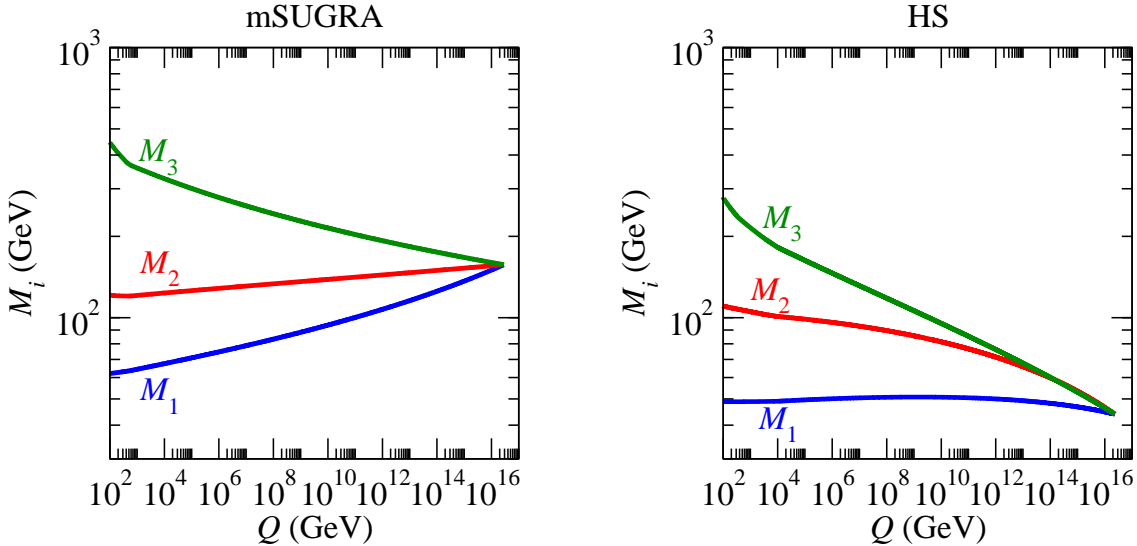}
\end{center}
\vspace*{-4mm}
\caption{\it 
Evolution of soft SUSY breaking gaugino mass parameters using two-loop RG evolution in the
case of mSUGRA and in the case of the HS model with parameters as listed in the text.
}\label{fig:inomass} 
\end{figure}

A more direct consequence of the Yukawa-unified SUSY models is that the color-octet
gluino particles are quite light, and possibly accessible to Fermilab Tevatron searches.
Under the assumption of gaugino mass unification, the LEP2 chargino mass limit that
$m_{\tw_1}>103.5$ GeV normally implies that $m_{\tg}\agt 430$ GeV, quite beyond the Tevatron
reach. However, in Yukawa-unified SUSY, the trilinear soft breaking term is large: 
$A_0\sim 10-20$ TeV.
Such a large trilinear term actually causes a large effect on gaugino mass 
evolution through two-loop RGE terms, as illustrated in Fig.~\ref{fig:inomass}. 
Here the left frame shows the gaugino mass evolution for the mSUGRA model point with 
$(m_0,m_{1/2},A_0,\tan\beta ,\mu)=(500\ {\rm GeV}, 157\ {\rm GeV},0,10,+)$, 
which has $m_{\tw_1}=103.5$ GeV, with $m_{\tg}=426.1$ GeV.
In the right frame, we show the gaugino mass evolution for Point B of Table 2 
of Ref. \cite{bhkss}, but with a slightly lower $m_{1/2}$ value. This point has the following
GUT scale input parameters: $m_{16}=10000$ GeV, $m_{10}=12053.5$ GeV, $M_D=3287.12$ GeV, 
$m_{1/2}=34$ GeV, $A_0=-19947.3$ GeV, $\tan\beta =50.398$ and $\mu >0$. 
($\tan\beta$ is input as a weak scale value.)
In this case,
the gaugino mass evolution is strongly affected by the large two-loop terms, 
resulting in a much smaller splitting between gaugino masses $M_2$ and $M_3$.
Here, we find (after computing physical masses including one-loop sparticle mass corrections) 
that  while $m_{\tw_1}=108.2$ GeV, the gluino mass is only $m_{\tg}=322.8$ GeV.
This value of $m_{\tg}$ may well be within range of Tevatron discovery, even while respecting
chargino mass bounds from LEP2.

In Yukawa-unified models, the $b$ and $\tau$ Yukawa couplings are large, while the top
and bottom squark masses are much lighter than their first/second generation counterparts.
As a consequence,
gluino decays to third generation particles-- in particular decays to $b$ quarks-- are
enhanced. 
In addition, gluino pair production via $q\bar{q}$ fusion is normally suppressed by
$t$- and $s$-channel interferences in the production cross section. For $m_{\tq}\sim 10$ TeV,
the negative interference is suppressed, leading to greatly enhanced gluino pair cross sections.
Use may be made of the large gluino pair production cross section, 
and the fact that each $\tg\tg$ production event is expected to have
four or more identifiable $b$-jets, along with large $\eslt$, to reject SM backgrounds.

In this letter, we examine gluino pair production at the Fermilab Tevatron collider.
While negative searches for gluino pair production have been made, and currently
require (under an analysis with $\sim 2$ fb$^{-1}$ of integrated luminosity) 
$m_{\tg}\agt 308$ GeV \cite{cdf,d0} in mSUGRA-like models, use has not yet been made 
of the large gluino pair production
cross section and high $b$-jet multiplicity expected
from Yukawa unified models.\footnote{The utility of $b$-jet tagging for extracting 
SUSY signals at the LHC has been examined in Ref's \cite{xt}.} 
Here, we point out the importance of exploiting the $b$-jet multiplicity to maximize the reach.
By requiring Tevatron events with $\ge$ 4 jets plus large $\eslt$, 
along with $\ge$ 2 or 3 tagged $b$-jets, QCD and electroweak 
backgrounds can be substantially reduced
relative to expected signal rates. We find that the CDF and D0 experiments should be
sensitive to $m_{\tg}\sim 400-440$ GeV with $5-10~{\rm fb}^{-1}$ of integrated luminosity.
Thus, Tevatron experiments are sensitive to much higher values of gluino mass than
otherwise expected from conventional searches.
With $5-10~{\rm fb}^{-1}$ of data, Tevatron experiments can indeed begin to explore a 
large swath of Yukawa-unified SUSY model parameter space.

In Sec.~\ref{sec:prod}, we review gluino pair production total cross sections and
expected branching fractions, and introduce a special Yukawa-unified SUSY model line.
In Sec.~\ref{sec:reach}, we provide details of our event simulation program, and show
how the requirement of events with $\ge 4$ jets plus large $\eslt$, along with $\ge 2$ or 3 
identified $b$-jets, rejects much SM background, at little cost to signal. We provide our
reach results versus $m_{\tg}$. In Sec.~\ref{sec:conclude}, we present a summary and conclusions. 

\section{Production and decay of gluinos at the Tevatron}
\label{sec:prod}

\subsection{Gluino pair production}

Recent studies of squark and gluino pair production at the Fermilab Tevatron collider, 
using data corresponding to $2$ fb$^{-1}$ of integrated luminosity and a beam energy
of $\sqrt{s}=1.96$ TeV, have produced limits at the 95\% CL that $m_{\tg}>280$ GeV
(in the case of CDF\cite{cdf}), and $m_{\tg}>308$ GeV (in the case of D0\cite{d0}).
These studies-- in the parts focused on gluino pair production-- essentially asked for the   
presence of events with $\ge 4$ hard jets, plus large $\eslt$ and large $H_T$, where $H_T$ is the scalar
sum of the $E_T$s of all identified jets in the event, beyond an expected SM background level.
These studies do not use some of the unique characteristics common to 
gluino pair production in Yukawa-unified SUSY, so we expect Tevatron experiments to be 
able to do much better in this case. 

\begin{figure}[p]
\begin{center}
\epsfig{file=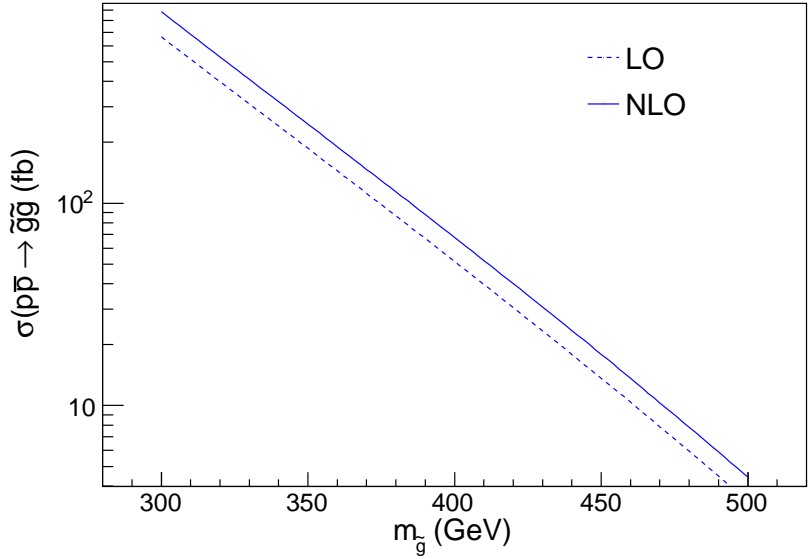,width=12cm}
\end{center}
\vspace*{-4mm}
\caption{\it 
Production cross section $\sigma (p\bar{p}\to\tg\tg X)$ in fb at the $\sqrt{s}=1.96$ TeV 
Fermilab Tevatron collider versus $m_{\tg}$, for $m_{\tq}=10$ TeV. 
}\label{fig:prod} 
\end{figure}

First, we present the expected total cross section rates for gluino pair production in
Fig.~\ref{fig:prod}, displaying leading order (LO) and next-to-leading order (NLO) 
cross sections as given by Prospino\cite{prospino}. We adopt a common first/second 
generation squark mass of $m_{\tq}=10$ TeV, and take the Tevatron energy as 
$\sqrt{s}=1.96$ TeV.  
We see from the figure that for $m_{\tg}=300$ GeV, the cross section is about 900~fb, 
dropping to about 65~fb for $m_{\tg}\simeq 400$ GeV. Moreover, it remains at the level 
of several fb even for $m_{\tg}$ as high as 500 GeV. 

\begin{figure}[p]
\begin{center}
\epsfig{file=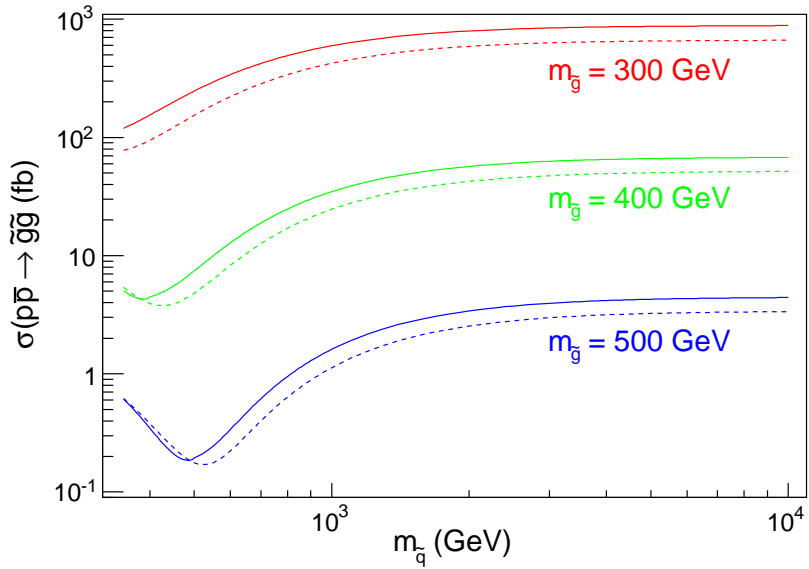,width=12cm}
\end{center}
\vspace*{-4mm}
\caption{\it 
Cross section of gluino pair production (in fb) at the Fermilab Tevatron collider 
as a function of $m_{\tq}$, for $m_{\tg}=300$, 400 and 500 GeV. Dashed is
LO QCD, while solid is NLO, as given by Prospino. 
}\label{fig:sigsq} 
\end{figure}

These cross sections are well in excess of those which enter the CDF and D0
search for gluino pair production. To understand why, we first note that gluino pair production
for $m_{\tg}\sim 300-500$ GeV is dominated by valence quark annihilation via 
$q\bar{q}$ fusion at the Tevatron. 
The $gg$ fusion subprocess is dominant at much lower gluino masses, 
where the gluon PDFs have their
peak magnitude at small parton fractional momentum $x$.
The $q\bar{q}\to\tg\tg$ subprocess cross section receives
contributions from $s$-channel gluon exchange, along with $t$- and $u$- channel squark 
exchange diagrams\cite{hl-s}. 
The $st$- and $su$-channel interference terms contribute {\it negatively}
to the total production cross section, thereby leading to an actual suppression of
$\sigma (p\bar{p}\to \tg\tg X)$ for $m_{\tq}\sim m_{\tg}$. 
For $m_{\tq}\gg m_{\tg}$ on the other hand, 
the $t$-channel, $u$-channel and interference terms are all highly 
suppressed, leaving the $s$-channel gluon exchange contribution unsuppressed and dominant. 
The situation is illustrated in Fig. \ref{fig:sigsq},
where we plot the LO and NLO gluino pair production cross section for $m_{\tg}=300$,
400 and 500 GeV versus $m_{\tq}$. 
We see that as $m_{\tq}$ grows, the total production cross section
{\it increases}, and by a large factor: for $m_{\tg}=400$ GeV, as $m_{\tq}$ varies from
400 GeV to 10 TeV, we see a factor of $\sim 10$ increase in total rate!

At the present time-- Fall 2009-- CDF and D0 have amassed over $5$ fb$^{-1}$ 
of integrated luminosity.\footnote{It is expected that CDF and D0 will reach the
$\sim 10$ fb$^{-1}$ level during 2010.}
Thus, if $m_{\tg}\sim 400$ GeV, there could be $\sim 300$ gluino pair events 
in each group's data.
Such a large event sample may well be visible if appropriate background rejection 
cuts can be found. 
The exact collider signatures depend on the dominant gluino decay modes, which we discuss in 
the next section.

\subsection{Gluino decays in Yukawa-unified SUSY}

To examine the gluino decay modes in Yukawa-unified SUSY, we will adopt a model line which allows
us to generate typical Yukawa-unified models over the entire range of $m_{\tg}$ which is expected.
First, we note in passing that Yukawa unification is not possible in the mSUGRA model, since the
large $t-b-\tau$ Yukawa couplings tend to drive the $m_{H_d}^2$ soft SUSY breaking term more negative
than $m_{H_u}^2$, in contradiction to what is needed for an appropriate breakdown of electroweak symmetry.
Yukawa-unified models can be found if one instead moves to models with non-universal Higgs masses,
where $m_{H_u}^2<m_{H_d}^2$ already at the GUT scale\cite{murayama,bdft}. 
In this case, $m_{H_u}^2$ gets a head start in its 
running towards negative values. Detailed scans over the parameter space in Ref. \cite{bkss} using 
the parameter space in \ref{eq:pspace} found a variety of solutions in the Higgs splitting (HS) model.
We will adopt Point~B of Table 2 of Ref. \cite{bhkss} as a template model. This point has the following
GUT scale input parameters: $m_{16}=10000$ GeV, $m_{10}=12053.5$ GeV, $M_D=3287.12$ GeV, 
$m_{1/2}=43.9442$ GeV, $A_0=-19947.3$ GeV, $\tan\beta =50.398$ and $\mu >0$,
(where $\tan\beta$ is again at the weak scale).
The Yukawa couplings at $M_{\rm GUT}$ are found to be
$f_t=0.557$, $f_b=0.557$ and $f_\tau =0.571$, so unification is good at the 2\% level.
The gluino mass which is generated is $m_{\tg}=351$ GeV. 

If we now allow $m_{1/2}$ to vary, we still maintain
valid Yukawa-unified solutions over the range of $m_{1/2}: 35- 100$ GeV, corresponding to a 
variation in $m_{\tg}: 325-508$ GeV. (The Yukawa unification gets worse as $m_{1/2}$ increases, 
and at $m_{1/2}=100$ GeV diminishes to 7.3\%.)
The value of the chargino mass at $m_{1/2}=35$ GeV is $m_{\tw_1}=108$ GeV, 
{\it i.e.}\ slightly above the LEP2 limit.
We will label Point~B with variable $m_{1/2}$ as the Higgs splitting, or HS, model line.
The value of the light Higgs boson is $m_h\simeq 127$ GeV all along the HS model line.   

\begin{figure}[!t]
\begin{center}
\epsfig{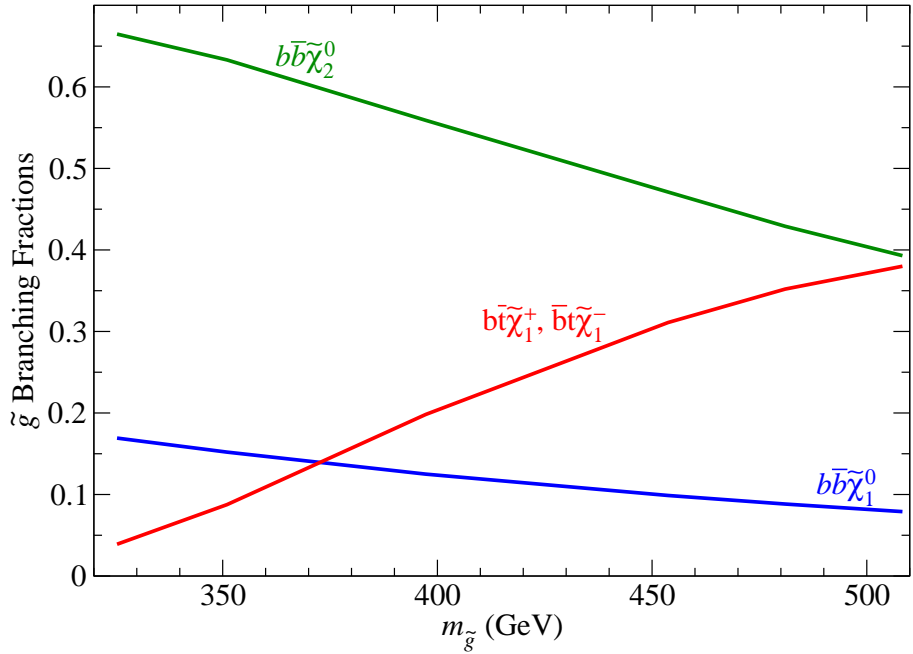}
\end{center}
\vspace*{-6mm}
\caption{\it 
Dominant gluino branching fractions versus $m_{\tg}$ for the 
Yukawa-unified SUSY HS model line.
}\label{fig:bf} 
\end{figure}

Armed with a Yukawa-unified SUSY model line, we can now examine how the gluino decays as a function
of gluino mass. The gluino decay branching fractions as calculated by Isajet are shown in Fig. \ref{fig:bf}.
Here, we see that at low $m_{\tg}\sim 325$ GeV, the mode $\tg\to b\bar{b}\tz_2$ occurs at over 60\%,
and dominates the $\tg\to b\bar{b}\tz_1$ branching fraction, which occurs at typically 10--20\%\cite{bcdpt}. 
As $m_{\tg}$ increases, the decay modes $\tg\to t\bar{b}\tilde{\chi}_1^- +c.c.$ grows from the kinematically
suppressed value of below 10\% at $m_{\tg}\sim 325$ GeV, to $\sim 40\%$ at $m_{\tg}\sim 500$ GeV.
All these dominant decay modes lead to two $b$s per gluino in the final state, so that for
gluino pair production at the Tevatron, we expect collider events containing almost always
$\ge 4$ jets $+\eslt$, with $\ge 4$ $b$-jets. Even more $b$-jets can come from $\tz_2$ decays, 
since $\tz_2\to b\bar{b}\tz_1$ at around 20\% all across the HS model line.
Only a very small fraction of gluino decays, less than 10\%,   
lead to first/second generation quarks in the final state.

\section{Reach of the Fermilab Tevatron for gluinos in Yukawa-unified SUSY}
\label{sec:reach}

Next, we examine whether experiments at the Fermilab Tevatron can detect
gluino pair production in the HS model line assuming 5-10 fb$^{-1}$ of integrated 
luminosity.  
We generate signal and background events
using Isajet 7.79, with a toy detector simulation containing hadronic
calorimetry ranging out to $|\eta | <4$, with cell size
$\Delta\eta\times\Delta\phi =0.1\times 0.262$.  We adopt hadronic
smearing of $\Delta E=0.7/\sqrt{E}$ and EM smearing of $\Delta
E=0.15/\sqrt{E}$. We adopt the Isajet GETJET jet finding algorithm,
requiring jets in a cone size of $\Delta R=0.5$ with $E_T^{{\rm jet}}>15$ GeV.
Jets are ordered from highest $E_T$ ($j_1$) to lowest $E_T$.  Leptons
within $|\eta_{\ell}| < 2.5$ ($\ell=e, \ \mu$) are classified as
isolated if $p_T(\ell )>10$ GeV and a cone of $\Delta R=0.4$ about the
lepton direction contains $E_T<2$~GeV. Finally,
if a jet with $|\eta_j|\le 2$ has a $B$-hadron with $E_T \ge 15$~GeV
within $\Delta R \le 0.5$, it is tagged as a $b$-jet with an efficiency
of 50\%. Ordinary QCD jets are mis-tagged as $b$-jets at a 0.4\% rate\cite{d0btag}.

We also generate SM background (BG) event samples from $W$ + jets production, 
$Z+b\bar{b}$ production, $t\bar{t}$ production, 
vector boson pair production, hadronic $b\bar{b}$ production, 
$b\bar{b}b\bar{b}$ production, $t\bar{t}b\bar{b}$ production and $Zb\bar{b}b\bar{b}$
(followed by $Z\to\nu\bar{\nu}$) production
\footnote{We do not take into account the QCD 
dijet backgrounds which turn out to be negligible after the cuts described below.}
The $W$ + jets sample uses QCD matrix elements for the primary parton
emission, while subsequent emissions (including $g\to b\bar{b}$ splitting) 
are generated from the parton shower.  
For $Z+b\bar{b}$, we use the exact $2\to 3$ matrix element, which is pre-programmed into Isajet
using Madgraph\cite{madgraph}.
We use AlpGen\cite{alpgen} plus Pythia\cite{pythia} for $b\bar{b}b\bar{b}$ and $t\bar{t}b\bar{b}$ production, and 
Madgraph plus Pythia for $Zb\bar{b}b\bar{b}$ production\cite{madgraph}.

\begin{figure}[!t]
\begin{center}
\epsfig{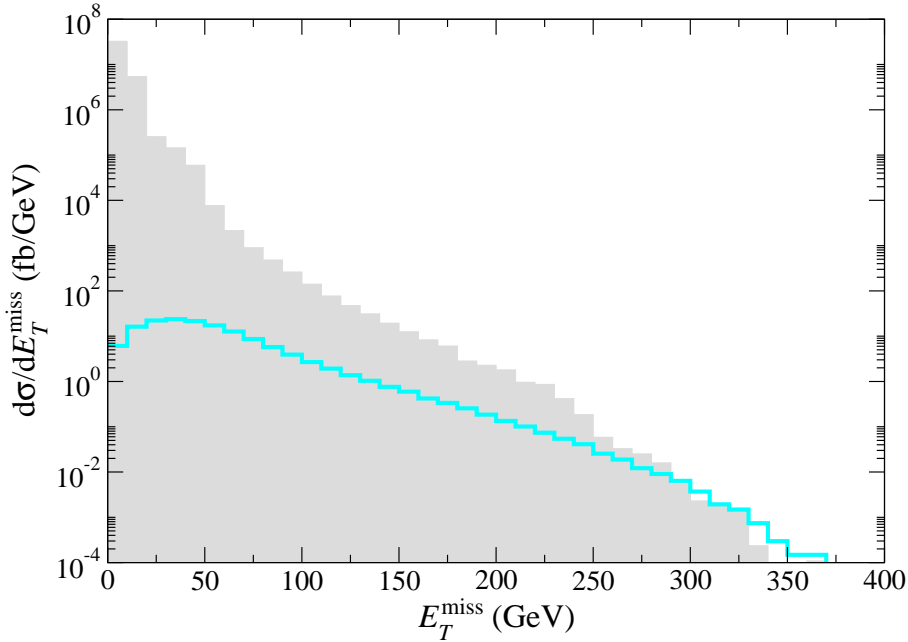}
\end{center}
\vspace*{-4mm}
\caption{\it 
Distribution in missing $E_T$ from gluino pair production in Yukawa-unified
SUSY Pt.~B, along with summed SM backgrounds (gray histogram), with minimal
cuts listed in text.
}\label{fig:etm} 
\end{figure}

For our first results, we exhibit the distribution in $\eslt$ in Fig. \ref{fig:etm} 
as generated for the HS model line Pt.~B (with $m_{\tg}=350$ GeV) as the blue histogram, along with
the summed SM backgrounds (gray histogram). 
While the signal histogram is harder than the BG histogram, 
the BG level is so high that signal doesn't exceed BG until $\eslt\agt 300$ GeV. 
Of course, this Pt.~B gluino mass is well beyond the current Tevatron gluino mass limits, 
so this is easy to understand.

To do better, we must adopt  a set of cuts that selects out canonical gluino pair production events.
Here, we will follow the recent papers BMPT\cite{bmpt}, CDF\cite{cdf} and D0\cite{d0}, 
and require the cuts listed in Table~\ref{tab:cuts}.\\

\begin{table}[h!]\centering
\begin{tabular}{lcccccc}
\hline
cuts & $\eslt$ & $H_T$ & $E_T(j1)$ & $E_T(j2)$ & $E_T(j3)$ & $E_T(j4)$ \\
\hline
BMPT & $\ge 75$ GeV & -- & 15 & 15 & 15 & 15 \\
CDF & $\ge 90$ GeV & 280 & 95 & 55 & 55 & 25 \\
D0 & $\ge 100$ GeV & 400 & 35 & 35 & 35 & 20 \\
\hline
\end{tabular}
\caption{\it Sets of cuts from Ref's \cite{bmpt}, \cite{cdf} and \cite{d0} used in this analysis. 
In addition we require throughout $\ge 4$ jets, {\it no} isolated leptons, 
at least one jet with $|\eta_j|<0.8$ and 
$\Delta\phi (j1,j2)<160^\circ$.
}\label{tab:cuts}
\end{table}

\begin{figure}[!t]
\begin{center}
\epsfig{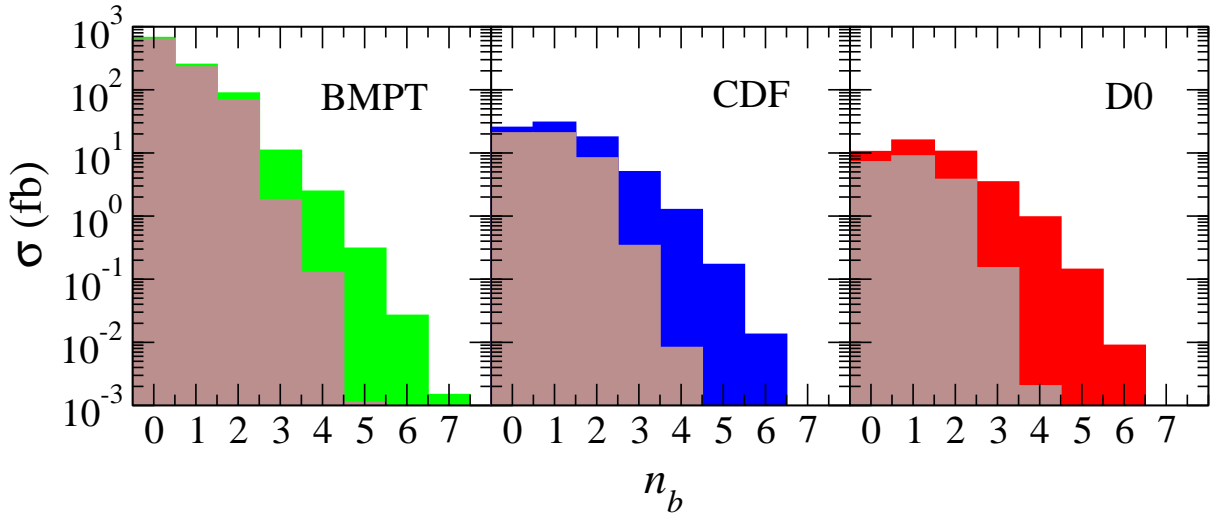}
\end{center}
\vspace*{-5mm}
\caption{\it 
Distribution in tagged $b$-jet multiplicity for gluino pair production in Yukawa-unified
SUSY Pt.~B, along with summed SM backgrounds (gray histogram), after 
cut sets BMPT, CDF, and D0 given in Table~\ref{tab:cuts} .
}\label{fig:nbj} 
\end{figure}

We have yet to make use of the high $b$-jet multiplicity which is expected
from Yukawa-unified SUSY. In Fig. \ref{fig:nbj}, we plot the multiplicity of $b$-jets
expected from SM background (brown histogram), and the summed BG plus signal from HS Pt.~B.
(The BG in the $n_b=0$ channel is very under-estimated, since we leave off QCD multi-jet production.)
We see that the BG distribution has a sharp drop-off as $n_b$ increases. 
Especially, there is a very sharp
drop-off in BG in going from the $n_b=2$ to the $n_b=3$ bin.
When we add in the signal distribution, we see the histogram expanding out to large
values of $n_b$ due to the presence of 4--6 $b$-jets per SUSY event. 
For the softer BMPT cuts, the signal hardly influences the
$n_b=0,1,2$ bins. 
However, in the $n_b=3$ bin, there is a huge jump in rate, reflecting the presence of a
strong source of $\ge 3$ $b$-jet events. 
In the case of the CDF and D0 cuts, which are much harder,
the total BG is much diminished. 
In this case, the summed signal plus BG distribution actually becomes rounded,
and is again much harder than just BG alone. For the CDF (D0) cuts, 
signal exceeds BG in the $n_b=2$ bin by a factor of 2 (3). By the time we move to the
$n_b=3$ bin, then for both CDF and D0 cuts, signal exceeds BG by over an order of magnitude.
Using soft cuts and low $b$-jet multiplicity, one should gain a good normalization of total BG rates.
Then, as one moves towards large $b$-jet multiplicity $n_b\ge 2$ or $3$, there should be much higher 
rates than expected from SM BG.

Table~\ref{tab:bg} shows a listing of expected contamination from each BG source after the different sets of cuts.
The hard $\eslt$ and $H_T$ cuts largely eliminate the $b\bar{b}$ BG. The isolated lepton veto 
combined with large $\eslt$ requirement cuts much of $W+jets$. The remaining large 
BGs come from $t\bar{t}$ production and $Z+b\bar{b}$ production, where $Z\to\nu\bar{\nu}$. 
Requiring $\ge 4$ jets along with large $H_T$ for the CDF and D0 cuts largely reduces
$Zb\bar{b}$ to small levels, leaving $t\bar{t}$ as the dominant BG.

\begin{table}\centering
\begin{tabular}{lcc|cc|cc|cc}
\hline
BG & $\sigma$ (fb) & events & \multicolumn{2}{|c|}{BMPT} & \multicolumn{2}{|c|}{CDF} & \multicolumn{2}{|c}{D0} \\
 & & & $\ge 2b$s & $\ge 3b$s & $\ge 2b$s & $\ge 3b$s & $\ge 2b$s & $\ge 3b$s \\ 
\hline
$b\bar{b}$ & $3.8\times 10^8$ & $10^6$ & $-$ & $-$ & $-$ & $-$ & $-$ & $-$ \\
$t\bar{t}$ & $5.9\times 10^3$ & $10^6$ & $51.9$ & $1.3$ & $8.6$ & $0.3$ & $3.9$ & $0.14$ \\
$b\bar{b}+(Z\to\nu\bar{\nu})$ & $1.3\times 10^4$ & $10^6$ & $15.7$ & $0.4$ & $-$ & $-$ & $-$ & $-$ \\
$W+jets$ & $4.8\times 10^6$ & $5\times 10^6$ & $1.9$ & $-$ & $-$ & $-$ & $-$ & $-$ \\
$VV$ & $9.7\times 10^3$ & $10^6$ & $0.6$ & $0.01$ & $-$ & $-$ & $-$ & $-$ \\ 
$b\bar{b}b\bar{b}$ & $6.3\times 10^4$ & $9.7\times 10^5$ & 0.39 & 0.13 & 0.065 & 0.065 & $-$ & $-$ \\
$t\bar{t}b\bar{b}$ & $11$ & $4.1\times 10^5$ & 0.39 & 0.13 & 0.066 & 0.019 & $0.037$ & $0.013$ \\
$b\bar{b}b\bar{b}+(Z\to\nu\bar{\nu})$ & $0.54$ & $6.6\times 10^3$ & $0.03$ & 
$0.01$ & $<10^{-2}$ & $<10^{-2}$ & $<10^{-3}$ & $<10^{-3}$ \\
\hline
Total & & & 70.9 & 1.98 & 8.7 & 0.38 & 3.94 & 0.15 \\
\hline
\end{tabular}\\[1mm]
\caption{\it 
SM backgrounds in fb before and after cuts BMPT, CDF and D0 for 
$n_b\ge 2$ and $\ge 3$. The $p_T$ range for $b\bar{b}$ subprocess generation is
$15-200$ GeV. The $p_T$ range for $t\bar{t}$ subprocess generation is $10-300$ GeV.
The $\sqrt{\hat{s}}$ range for $Zb\bar{b}$ subprocess generation is $100-400$ GeV.
In the above, $V=W$ or $Z$.
}\label{tab:bg}
\end{table}

\begin{figure}[!t]
\begin{center}
\epsfig{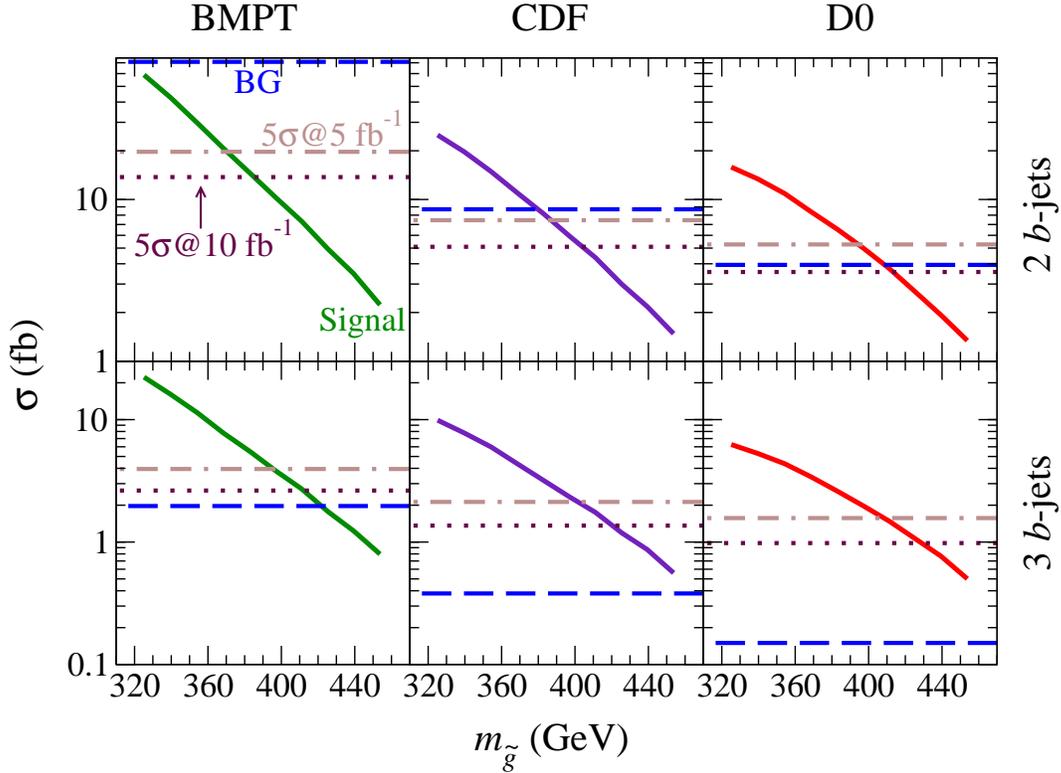}
\end{center}
\vspace*{-4mm}
\caption{\it 
Reach of the Fermilab Tevatron collider for gluino pair production in 
Yukawa-unified SUSY HS model line, versus $m_{\tg}$. We show the
reach for 5 and 10 fb$^{-1}$ of integrated luminosity. 
}\label{fig:reach} 
\end{figure}

In light of these results, we proceed by requiring BMPT, CDF or D0 cuts, along with
\bi
\item $n_b \ge 2$ or 3.
\ei
In Fig. \ref{fig:reach} we plot the resultant SM background (blue dashed lines),
along with expected signal rates for the HS model line (full lines) for the three 
sets of cuts with $n_b \ge 2$ (upper row) as well as $n_b \ge 3$ (lower row). 
The SM background comes
almost entirely from $t\bar{t}$ production. The third $b$-jet in $t\bar{t}$ 
production can come from additional $g\to b\bar{b}$ radiation, or from QCD jet mis-tags.
Since the dominant BG comes from $t\bar{t}$ production, and the $\sigma (p\bar{p}\to t\bar{t}X )$
cross section is well-known from standard top search channels, the background should be rather well
understood.

We see from Fig. \ref{fig:reach} that signal actually exceeds BG for a substantial
range of $m_{\tg}$ for all cases except the BMPT cuts with $n_b\ge 2$. 
We also compute the signal cross sections required for a $5\sigma$ discovery for each selection 
assuming 5 and 10 fb$^{-1}$ of integrated luminosity, shown as the dot-dashed and dotted lines, 
respectively.  The significance in $\sigma$s is derived from the p-value corresponding to the number of S+B 
events in a Poisson distribution with a mean that equals to the number of background events.
The best reach is achieved with the hard D0 cuts. In this case, requiring $n_b\ge 2$, 
we find that signal exceeds the $5\sigma$ level for 
5 (10) fb$^{-1}$ of integrated luminosity for $m_{\tg}<395\ (410)$ GeV. 
Requiring $n_b\ge 3$, the $5\sigma$ reach for 5 (10) fb$^{-1}$ increases to 
$m_{\tg}=405$ (430) GeV.
Thus, in the case of Yukawa-unified SUSY where an abundance of $b$-jets
are expected to accompany gluino pair production, 
{\it we expect Fermilab Tevatron experiments to be able to probe values of 
$m_{\tg}$ to much higher values than have previously been found.}

Since the value of $m_{\tg}$ is expected to lie in the range 300--500 GeV for Yukawa-unified models,
and in fact the Yukawa unification is best on the lower range of $m_{\tg}$ values, it appears
to us that CDF and D0, using current data samples, stand a good chance of either discovering
Yukawa-unified SUSY, or excluding a huge portion of the allowed parameter space.

\section{Conclusions}
\label{sec:conclude}

In this paper, we explored the capability of the
CDF and D0 experiments to search for gluinos with properties as
predicted by supersymmetic models with $t-b-\tau$ Yukawa coupling unification.
While a vast effort is rightfully being placed by CDF and D0 to search for
the SM Higgs boson, a potentially bigger prize-- the gluinos from supersymmetric models--
could be lurking in their data. The Yukawa-unified SUSY model is extremely compelling, 
in part because it combines four of the most profound ideas in physics beyond the SM: 
$SO(10)$ grand unification (which unifies matter as well as gauge couplings),
weak scale supersymmetry, see-saw neutrino masses and the Peccei-Quinn-Weinberg-Wilczek
solution to the strong CP problem.
While we do not present a specific model which incorporates all these ideas 
into a single framework,
a wide array of low energy, collider and astrophysical data give some indirect and also direct support
to each of these ideas. The requirement of Yukawa coupling unification forces upon us a 
very specific and compelling sparticle mass spectrum, including first/second generation scalars at the
$\sim 10$ TeV scale, while gluinos are quite light: in the $\sim$ 300--500 GeV range.
We investigated here whether these light gluinos are accessible to Tevatron searches 
for supersymmetry.

Our main result is that the CDF and D0 experiments should be already sensitive to gluino masses
far beyond currently published bounds (which lie around the 300 GeV scale). This is due to
three main factors: 
\begin{enumerate}
\item Two-loop RGE effects allow for gluinos as light as 320 GeV in the Yukawa-unified model
with multi-TeV trilinear soft terms, even while respecting LEP2 limits on the chargino mass.
In the case of generic SUSY models with TeV scale soft parameters, the LEP2 chargino mass limit
usually implies $m_{\tg}\agt 425$ GeV.
\item Gluino pair production cross sections with $m_{\tg}\sim 300-500$ GeV 
are enhanced at the Tevatron due to the extremely high squark masses expected in Yukawa-unified SUSY.
The huge value of $m_{\tq}$ acts to suppress negative interference effects in the 
$q\bar{q}\to\tg\tg$ subprocess cross section, leading to elevated production rates.
\item Gluinos of Yukawa-unified SUSY decay through cascade decays to final states almost
always containing four $b$-jets, and sometimes six or eight $b$-jets, depending if $\tz_2\to \tz_1b\bar{b}$
occurs.
By searching for collider events with $\ge 4$ jets plus large $\eslt$, along with $\ge 2$ or 3 
$b$-jets which are tagged through the micro-vertex detector, SM backgrounds can be reduced by
large factors, at only a small cost to signal. 
\end{enumerate}
This may allow Tevatron experiments to search for gluinos with mass in excess of 400 GeV. 
Such gluino masses are far beyond currently published bounds,
and would allow exploration of a huge swath of parameter space of Yukawa-unified SUSY models.

In addition, in the case of the HS model where $\tg\to b\bar{b}\tz_2$ at a large rate, followed
by $\tz_2\to\tz_1 \ell^+\ell^-$ (typically at $\sim 3\%$ branching ratio for each of $\ell =e$ or $\mu$),
there may be a corroborating signal at much lower rates in the multi-$b$-jet$+\eslt +\ell^+\ell^-$ mode,
where $m(\ell^+\ell^- )<m_{\tz_2}-m_{\tz_1}$.

We note finally that the results presented here in the context of Yukawa-unified models
are more generally applicable to any model with very heavy scalars, and large enough
$\tan\beta$ such that gluinos dominantly decay via three-body modes into $b$-quarks.
They are also applicable to models with hierarchical soft terms, where first/second generation
scalars are extremely heavy, and third generation scalars are much lighter; some references for such models
are located in \cite{hst}.

\section*{Acknowledgments} 
We thank Phil Gutierrez for a discussion and Xerxes Tata for comments 
on the manuscript. We thank Harrison Prosper for discussion on 
statistics issues and sharing with us his p-value and significance 
code.  This work is funded in part by the US Department of Energy, 
grant number DE-FG-97ER41022 and by the French ANR project 
ToolsDMColl, BLAN07-2-194882.



\begin{thebibliography}{99}
\small
%
%
\bibitem{wss} H.~Baer and X.~Tata, {\it Weak Scale Supersymmetry: From 
Superfields to Scattering Events}, 
(Cambridge University Press, 2006).
%
\bibitem{cdf} T. Aaltonen {\it et al.} (CDF Collaboration), \prl{102}{2009}{121801}. 
%
\bibitem{d0} V. M. Abazov {\it et al.} (D0 Collaboration), 
\plb{660}{2008}{449}.
%
\bibitem{raby} For recent reviews,   
see R. Mohapatra, hep-ph/9911272 (1999) and S. Raby, in
Rept. Prog. Phys. {\bf 67} (2004) 755.
%
\bibitem{abbbft} D. Auto, H. Baer, C. Balazs, A. Belyaev, J. Ferrandis 
and X. Tata, \jhep{0306}{2003}{023}.
%
\bibitem{bkss} H. Baer, S. Kraml, S. Sekmen and H. Summy,
\jhep{0803}{2008}{056}.
%
\bibitem{agrs} W. Altmannshofer, D. Guadagnoli, S. Raby and D. Straub, \plb{668}{2008}{385}.
%
\bibitem{bhkss} H. Baer, M. Haider, S. Kraml,  S. Sekmen and H. Summy,
JCAP{\bf 0902} (2009) 002.
%
\bibitem{dr3} H. Baer, S. Kraml and S. Sekmen, \jhep{0909}{2009}{005}.
%
\bibitem{so10} H. Georgi, in {\it Proceedings of the American Institue   
of Physics}, edited by C. Carlson (1974); H. Fritzsch and P. Minkowski,  
Ann. Phys. {\bf 93}, 193 (1975); M. Gell-Mann, P. Ramond and R. Slansky,  
Rev. Mod. Phys. {\bf 50}, 721 (1978). 
%
\bibitem{seesaw}  
P. Minkowski, \plb{67}{1977}{421};
M. Gell-Mann, P. Ramond and R. Slansky, 
in {\it Supergravity,  
Proceedings of the Workshop}, Stony Brook, NY 1979 (North-Holland, 
Amsterdam);  
T. Yanagida, KEK Report No. 79-18, 1979; R. Mohapatra and G. Senjanovic,   
\prl{44}{1980}{912}.
%
\bibitem{early} B. Ananthanarayan, G.~Lazarides and Q.~Shafi, 
\prd{44}{1991}{1613} and \plb{300}{1993}{245}; 
G.~Anderson {\it et al.} \prd{47}{1993}{3702} and \prd{49}{1994}{3660};
V. Barger, M. Berger and P. Ohmann,   
\prd{49}{1994}{4908};
M. Carena, M. Olechowski, S. Pokorski and C. Wagner,  
Ref. \cite{hrs};   
B. Ananthanarayan, Q. Shafi and X. Wang, \prd{50}{1994}{5980};
R. Rattazzi and U. Sarid, \prd{53}{1996}{1553};
T.~Blazek, M.~Carena, S.~Raby and C.~Wagner, \prd{56}{1997}{6919}; 
T.~Blazek and S. Raby, \plb{392}{1997}{371};
T.~Blazek and S.~Raby, \prd{59}{1999}{095002};
T.~Blazek, S.~Raby and K.~Tobe, \prd{60}{1999}{113001}
and \prd{62}{2000}{055001}; see also \cite{also}.
%
\bibitem{also} S. Profumo, \prd{68}{2003}{015006}; C. Pallis, \npb{678}{2004}{398};
M. Gomez, G. Lazarides and C. Pallis, 
\prd{61}{2000}{123512}, \npb{638}{2002}{165} and \prd{67}{2003}{097701};
U. Chattopadhyay, A. Corsetti and P. Nath, \prd{66}{2002}{035003};
M. Gomez, T. Ibrahim, P. Nath and S. Skadhauge,
\prd{72}{2005}{095008}.
%
\bibitem{dr} R. Dermisek and S. Raby, \prd{62}{2000}{015007};
R. Dermisek, M. Harada and S. Raby, \prd{74}{2006}{035011}. 
%
\bibitem{xdguts}Y. Kawamura, \ptp{105}{2001}{999}; 
G. Altarelli and F. Feruglio, \plb{511}{2001}{257}; 
L. Hall and Y. Nomura, \prd{64}{2001}{055003};  
A. Hebecker and J. March-Russell, \npb{613}{2001}{3};
A. Kobakhidze, \plb{514}{2001}{131}.
%
\bibitem{hrs} R.~Hempfling, \prd{49}{1994}{6168};  
L. J. Hall, R. Rattazzi and U. Sarid,   
\prd{50}{1994}{7048}; M.~Carena {\it et al.},  
\npb{426}{1994}{269};
D. Pierce, J. Bagger, K. Matchev and R. Zhang, \npb{491}{1997}{3}.
%
\bibitem{isajet} ISAJET v7.79, by H. Baer, F. Paige, S. Protopopescu and
X. Tata, \hepph{0312045}; for details on the Isajet spectrum calculation, see
H. Baer, J. Ferrandis, S. Kraml and W. Porod, \prd{73}{2006}{015010}.
%
\bibitem{bdft} H. Baer, M. Diaz, J. Ferrandis and X. Tata, 
\prd{61}{2000}{111701}; H. Baer, M. Brhlik, M. Diaz, J. Ferrandis,
P. Mercadante, P. Quintana and X. Tata, \prd{63}{2001}{015007}.
%
\bibitem{bf} H. Baer and J. Ferrandis, \prl{87}{2001}{211803};
%
\bibitem{bdr} T. Blazek, R. Dermisek and S. Raby, \prl{88}{2002}{111804};
T. Blazek, R. Dermisek and S. Raby, \prd{65}{2002}{115004};
R. Dermisek, S. Raby, L. Roszkowski and
R. Ruiz de Austri, \jhep{0304}{2003}{037};
R. Dermisek, S. Raby, L. Roszkowski and
R. Ruiz de Austri, \jhep{0509}{2005}{029}.
%
\bibitem{mtop} The Tevatron Electroweak Working group (CDF and D0 Collaboration\
s), arXiv:0803.1683.
%
\bibitem{slha}
  P.~Skands {\it et al.}, \jhep{0407}{2004}{036}
%
\bibitem{bfpz} J. Feng, C. Kolda and N. Polonsky, \npb{546}{1999}{3}; 
J. Bagger, J. Feng and N. Polonsky, \npb{563}{1999}{3};
J. Bagger, J. Feng, N. Polonsky and R. Zhang, \plb{473}{2000}{264}.
H. Baer,P. Mercadante and X. Tata, \plb{475}{2000}{289};
H. Baer, C. Balazs, M. Brhlik, P. Mercadante, X. Tata and Y. Wang, \prd{64}{2001}{015002}.
%
\bibitem{splita}
  D.~Guadagnoli, S.~Raby and D.~M.~Straub, arXiv:0907.4709 [hep-ph].
%
\bibitem{wmap} 
J.~Dunkley {\it et al.}  [WMAP Collaboration], Astrophys.\ J.\ Suppl.\  {\bf 180}, 306 (2009)
%
\bibitem{abbk} D. Auto, H. Baer, A. Belyaev and T. Krupovnickas, \jhep{0410}{2004}{066}. 
%
\bibitem{nr} H. P. Nilles and S. Raby, \npb{198}{1982}{102};
J. E. Kim and H. P. Nilles, \plb{138}{1984}{150};
J. E. Kim, \plb{136}{1984}{378}.
%
\bibitem{ntlepto} G. Lazarides and Q. Shafi, \plb{258}{1991}{305};
K. Kumekawa, T. Moroi and T. Yanagida, \ptp{92}{1994}{437};
T. Asaka, K. Hamaguchi, M. Kawasaki and T. Yanagida, \plb{464}{1999}{12}.
%
\bibitem{admx} L. Duffy {\it et al.}, \prl{95}{2005}{091304} and \prd{74}{2006}{012006};
for a review, see S. Asztalos, L. Rosenberg, K. van Bibber, P. Sikivie
and K. Zioutas, \arnps{56}{2006}{293}.
%
\bibitem{xt} U. Chattoppadhyay, A. Datta, A. Datta, A. Datta and D.P. Roy, 
\plb{493}{2000}{127}; P.~Mercadante, K.~Mizukohi and X. Tata,
\prd{72}{30005}{035009}; S. P. Das et al. \epjc {54}{2008}{645};
R. Kadala et al. \epjc {56}{2008}{511}
%
\bibitem{prospino} W. Beenakker, R. Hopker, M. Spira, \hepph{9611232} (1996).
%
\bibitem{hl-s} P. R. Harrison and C. H. Llewellyn-Smith, \npb{213}{1983}{223}.
%
\bibitem{murayama} H. Murayama, M. Olechowski and S. Pokorski,
\plb{371}{1996}{57}.
%
\bibitem{bcdpt} H. Baer, C. H. Chen, M. Drees, F. Paige and X. Tata, 
\prl{79}{1997}{986} and \prd{58}{1998}{075008}.
%
\bibitem{d0btag} V. M. Abazov {\it et al.} (D0 Collaboration), \plb{626}{2005}{35}.
%
\bibitem{madgraph} T. Stelzer and W. F. Long, \cpc{81}{1994}{357}; 
F. Maltoni and T. Stelzer, \jhep{0302}{2003}{027};
J. Alwall {\it et al.}, \jhep{0709}{2007}{028}.
%
\bibitem{alpgen}  M. Mangano, M. Moretti, F. Piccinini, R. Pittau and
A. Polosa, \jhep{0307}{2003}{001}.
%
\bibitem{pythia} T. Sjostrand, S. Mrenna and P. Skands,
\jhep{0605}{2006}{026}.
%
\bibitem{bmpt} H. Baer, A. Mustafayev, S. Profumo and X. Tata,
\prd{75}{2007}{035004}.
%
\bibitem{hst} A. Cohen, D. B. Kaplan and A. Nelson, \plb{388}{1996}{588};
H. Baer, C. Balazs, P. Mercadante, X. Tata and Y. Wang, \prd{63}{2001}{015011};
G. Giudice, M. Nardecchia and A. Romanino, \npb{813}{2009}{156}.
%
\end{thebibliography}
\end{document}